\documentclass[aps,pre,reprint,groupedaddress]{revtex4-1}

\usepackage{amsmath}
\usepackage{amssymb}
\usepackage{graphicx}
\usepackage{xcolor}

\newcommand{\dif}{d}

\newcommand{\idif}[2][\,]{\,\dif^{#1\!}#2}
\newcommand{\bigO}{\mathrm O}
\newcommand{\deBroglie}{\Lambda}
\newcommand{\mf}{\mathrm{mf}}
\newcommand{\scr}{\mathrm{scr}}

\renewcommand{\vec}[1]{\boldsymbol{#1}}
\newcommand{\mat}[1]{\mathcal{#1}}

\let\Im\relax\DeclareMathOperator\Im{Im}

\begin{document}


\title{Correlations Between Conduction Electrons in Dense Plasmas}


\author{Nathaniel R. Shaffer}
\email[]{nshaffer@lanl.gov}
\affiliation{Los Alamos National Laboratory}

\author{Charles E. Starrett}
\affiliation{Los Alamos National Laboratory}


\date{\today}

\begin{abstract}
  Most treatments of electron-electron correlations in dense plasmas either ignore them entirely (random phase approximation) or neglect the role of ions (jellium approximation).
  In this work, we go beyond both these approximations to derive a new formula for the electron-electron static structure factor which properly accounts for the contributions of both ionic structure and quantum-mechanical dynamic response in the electrons.
  The result can be viewed as a natural extension of the quantum Ornstein-Zernike theory of ionic and electronic correlations, and it is suitable for dense plasmas in which the ions are classical and the conduction electrons are quantum-mechanical.
  The corresponding electron-electron pair distribution functions are compared with the results of path integral Monte Carlo simulations, showing good agreement whenever no strong electron resonance states are present.
  We construct approximate potentials of mean force which describe the effective screened interaction between electrons.
  Significant deviations from Debye-H\"uckel screening are present at temperatures and densities relevant to high energy density experiments involving warm and hot dense plasmas.
  The presence of correlations between conduction electrons is likely to influence the electron-electron contribution to the electrical and thermal conductivity.
  It is expected that excitation processes involving the conduction electrons (e.g., free-free absorption) will also be affected.
\end{abstract}


\maketitle

\section{Introduction}
\label{sec:intro}

In a simple description of metals and plasmas, the conduction electrons may be regarded as weakly interacting because their kinetic energy is large compared to their mutual Coulomb repulsion.
Such is the case in the limits of both low and high temperature, where the respective kinetic energy scales are the Fermi energy and the temperature.
Electron transport at each extreme is modeled well by the Ziman theory of liquid metals or the Spitzer-H\"arm theory of classical plasmas, respectively\cite{ZimanPM1961,SpitzerPR1953}.
Warm and hot dense plasmas occupy an intermediate regime where the Fermi energy and temperature are of similar order, typically occurring at temperatures from a few eV to a few keV and mass densities ranging from fractions of solid density to hundreds of times solid density. 
In the laboratory, such conditions occur in inertial confinement fusion implosions\cite{HuPRL2010,GaffneyHEDP2018,ZaghooPRL2019}, in exploding wire arrays\cite{BenagePOP2000}, and in pulse power devices\cite{ClerouinPOP2012,NagayamaPRL2019}.
In Nature, one finds partially degenerate plasmas in the envelopes of white dwarfs and in the solar interior\cite{FontainePASP2001,SalpeterApJ1969}.
It is in this regime that the conduction electrons may develop significant spatial correlations with one another, and these correlations will impact electron transport and optical processes.

The need for new theoretical descriptions of electron-electron correlations in dense plasmas has been brought to light by recent work highlighting the importance of electron-electron scattering on electrical and thermal conduction in partially degenerate plasmas\cite{ReinholzPRE2012,ReinholzPRE2015,DesjarlaisPRE2017,DuftyCPP2018}.
Such conditions are challenging for quantum simulation methods, the most widespread being density functional theory molecular dynamics paired with the Kubo-Greenwood method for electron transport\cite{HuPRE2014b,LambertPOP2011,HolstPRB2011,HuPOP2016,DesjarlaisPRE2017}.
These simulations scale poorly with increasing temperature, and the use of the Kubo-Greenwood method introduces an approximate treatment of electron-electron scattering\cite{DesjarlaisPRE2017,DuftyCPP2018}.
It is not yet fully understood to what degree the Kubo-Greenwood approximation affects QMD predictions of transport properties, especially thermal conductivity.
This means that currently there is a wide span in temperatures between warm dense matter conditions and classical plasma conditions where quantum simulations are impractical and possibly inaccurate, yet the influence of correlations on electron-electron scattering is likely to affect transport in ways that classical plasma theory cannot predict.

While electronic correlation in metals has been an active area in condensed matter physics for decades, many theoretical developments in that field do not transfer in an obvious way to plasmas, where the high temperatures mean that the ions are not arranged on a lattice and the Fermi surface is not an especially useful construct to understand the electron dynamics.
For this reason, theoretical treatments of electron-electron correlations in dense plasmas commonly adopt the random phase approximation (in which electron correlations are ignored) and/or the jellium approximation (in which the electron correlation properties are co-opted from those of the homogeneous electron gas).
More sophisticated approaches based on the Green's function formalism have also been explored\cite{ReinholzPRE2000,ReinholzPRE2012}.
The limited knowledge of electron-electron correlations in plasma also affects experiments, since models of the plasma dynamic structure factor are used to diagnose the plasma density and temperature from x-ray diagnostics\cite{GlenzerRMP2009,CrowleyHEDP2014,BaczewskiPRL2016}.

This work provides, to our knowledge, the first accurate account of static correlations between the conduction electrons of dense plasmas.
The main result is a new expression for the electron-electron static structure factor appropriate for dense plasmas, which goes beyond the widely used random phase and jellium approximations by accounting both for direct correlations between the electrons as well as indirect correlations by the surrounding ions.
The focus here is mainly on static electron-electron correlations; however, this already should serve as a useful starting point for building theories of dynamic correlations in dense plasma in the adiabatic approximation or in a generalized dynamic linear response formalism\cite{ReinholzPRE2012,ReinholzPRE2015}.
Our results should also be useful in formulating new approximations to the electron self-energy via the inverse dielectric function, thereby facilitating the application of Green's function techniques such as $GW$ to study free-free excitations in dense plasma\cite{AryasetiawanRMP1998}.
Similarly, our results would be of use in constructing new exchange-correlation functionals that accurately treat the free electrons of dense plasmas within density functional theory\cite{MartinElectronicStructure}, or new adiabatic approximations to the exchange-correlation kernel for time-dependent density functional theory\cite{MarquesTDDFT,DuftyPRE2018}.
Specifically, the electron-electron correlation functions predicted here contain the ionic correlations explicitly, which are not directly accounted for in exchange-correlation functionals based on jellium.
Developments along these lines would also have applications to predicting the influence of electron correlations on photo-excitation processes involving conduction electrons, e.g., free-free absorption\cite{ReinholzPRE2012,HuPRE2014a,ShafferHEDP2017,HollebonPRE2019}

The expression for the structure factor derived here differs from the result one would obtain classically by the appearance of a term which accounts for quantum-mechanical dynamic screening.
The result follows from general linear response considerations and naturally extends the quantum Ornstein-Zernike theory of ion-ion and electron-ion correlations\cite{ChiharaPTP1973a,AntaPRB2000,StarrettPRE2013}.
When suitably paired with an average-atom treatment of electronic structure, the quantum Ornstein-Zernike relations are known to give a realistic description of both the ionic and electronic structure of dense plasmas\cite{StarrettPRE2013}.
With mild approximations, our result for the electron-electron structure factor is cast in a form that is amenable to practical calculations with average-atom models.
From this, we compute the pair distribution functions of warm/hot dense hydrogen and aluminum and compare with available path integral Monte Carlo results on fully ionized plasmas, finding good agreement when the notion of ``free'' and ``bound'' electrons in the average-atom model is well-defined, e.g., when there are no long-lived resonance states.
We also construct an approximate electron-electron potential of mean force and contrast it with the high-temperature limit where the plasma is weakly coupled and the effective potential is described well by exponential Debye-H\"uckel screening\cite{DebyePZ1923}.
Mean-force potentials are a promising means of modeling electron correlations' effect on the transport properties of dense plasmas within the framework of binary-scattering kinetic theories\cite{BaalrudPRL2013,DaligaultPRL2016,StarrettHEDP2017}.
In such a model, the electron-electron mean-force potential would improve on Spitzer and H\"arm's treatment of electron-electron scattering at dense plasma conditions within the static screening approximation.
At lower temperatures, significant deviations from exponential screening are observed and attributed both to indirect correlations induced by the strongly coupled ions as well as core-valence orthogonality.

\section{Theory}
\label{sec:theory}

\subsection{Quantum Ornstein-Zernike Description of a Two-Component Plasma}
\label{sec:tcp}

We model a dense plasma as a two-component mixture of classical point ions with mean number density $n_I^0$ and conduction electrons with mean number density $\bar n_e^0$.
The plasma is assumed neutral so that the mean degree of ionization is $\bar Z = \bar n_e^0/n_I^0$, which is density- and temperature dependent and may be fractional.
In this work, the ionization and thus the electron density are obtained from the average-atom two-component plasma (AA-TCP) model\cite{StarrettPRE2013}.
The notation adopted for densities and ionization is chosen to match Ref.~\cite{StarrettPRE2013}.

The central equations governing the AA-TCP model are the quantum Ornstein-Zernike (QOZ) equations.
These express the static structure factors of the TCP, $S_{ab}(k)$, in terms of the unknown direct correlation functions, $C_{ab}(k)$,
\begin{subequations}
  \label{eq:qoz}
  \begin{align}
    \label{eq:qoz-ii}
    & S_{II}(k) = \frac{1+\beta^{-1}\chi_e^0(k)C_{ee}(k)}{D(k)}
    \\
    \label{eq:qoz-ie}
    & S_{Ie}(k) = \bar Z^{-\frac12} n_e^\scr(k) S_{II}(k)
    \\
    \label{eq:qoz-nescr}
    & n_e^\scr(k) = \frac{-\beta^{-1}\chi_e^0(k) C_{Ie}(k)}{1 +\beta^{-1}\chi_e^0(k)C_{ee}(k)}
    \\
    \label{eq:qoz-d}
    & D(k) = (1 - n_I^0 C_{II})(1 - \beta^{-1}\chi_e^0  C_{ee}) - n_I^0\beta^{-1}\chi_e^0 |C_{Ie}|^2
  \end{align}
\end{subequations}
where $\chi_e^0(k)$ is the static density response function of noninteracting electrons, which is equal to $-\bar n_e^0\beta$ in the classical limit and is the Lindhard function at zero temperature.
The solution of the QOZ equations for $S_{II}(k)$ and $S_{Ie}(k)$ requires closure relations for the direct correlation functions $C_{II}$, $C_{Ie}$, and $C_{ee}$.
These closures complete the AA-TCP model.
The specific closures used in this work are described in the Appendix.

Observe that in the QOZ equations, Eq.~\eqref{eq:qoz}, no expression is given for the electron-electron structure factor, $S_{ee}(k)$.
In the literature on the QOZ theory, one can find equations for the electron-electron zero-frequency susceptibility, $\chi_{ee}(k,\omega=0)$\cite{ChiharaJPCM1984,AntaPRB2000,StarrettPRE2013}.
However, such formulas are unsuitable for describing the electron-electron static structure.
This is because electron-electron correlations must be treated quantum-mechanically.
In the quantum theory of correlation functions, the static limit and the zero-frequency limits are \emph{not} equivalent, in marked contrast to the classical case\cite{IchimaruBook}.
A consequence is that the calculation of $S_{ee}(k)$ -- despite being a static correlation function -- still requires accounting for the quantum-mechanical dynamic response of electrons.
Sec.~\ref{sec:lin-rep} will demonstrate this from completely general linear response considerations.
Then, with some mild assumptions, an extended set of QOZ equations are derived which include a relation for $S_{ee}(k)$ that is correct quantum-mechanically.

\subsection{Linear Response and Extended QOZ Relations}
\label{sec:lin-rep}

The dynamic density-density response functions for a multi-species plasma obey\cite{IchimaruBook,QSCPSBook,ReinholzAPF2005}
\begin{equation}
  \mat X = \mat X^0 + \mat X^0  \mat U \mat X
\end{equation}
where $\mat X$ is the matrix of response functions $\chi_{ab}(k,\omega)$, $\mat X^0$ is the matrix of free-particle response functions $\chi^0_{a}(k,\omega)\delta_{ab}$, and $\mat U$ is the matrix of polarization potentials $U_{ab}(k,\omega) = v_{ab}(k)[1 - G_{ab}(k,\omega)]$ expressed in terms of the Coulomb interaction $v_{ab}(k)=4\pi Z_aZ_be^2/k^2$ and the dynamic local field corrections $G_{ab}(k,\omega)$.
For a TCP, we can explicitly solve for the response functions
\begin{subequations}
  \label{eq:lin-resp}
  \begin{align}
    & \chi_{II}(k,\omega) = \chi_I^0(k,\omega)\frac{1 - \chi_e^0(k,\omega)U_{ee}(k,\omega)}{D(k,\omega)} \\
    & \chi_{Ie}(k,\omega) = -\chi_I^0(k,\omega)\frac{\chi_e^0(k,\omega)U_{Ie}(k)}{D(k,\omega)} \\
    & \chi_{ee}(k,\omega) = \chi_e^0(k,\omega)\frac{1 - \chi_I^0(k,\omega)U_{II}(k,\omega)}{D(k,\omega)} \\
    & D(k,\omega) = \det\left\{ \delta_{ab} - \chi_a^0(k,\omega)U_{ab}(k,\omega)  \right\}
      ,
  \end{align}
\end{subequations}
Taking all species to be fermions\footnote{
  In the analysis of Sec.~\ref{sec:lin-rep}, we treat all species as fermions.
  This is not necessarily true of the ions, but the distinction does not matter once the classical limit is taken.
  The same final results would be obtained assuming the ions were bosons.
}, the free-particle response functions are given by
\begin{equation}
  \label{eq:chi0}
  \chi_a^0(k,\omega) = -\beta n_a I_a(k,\omega)
\end{equation}
with\cite{TanakaJPSJ1986}
\begin{multline}
  \label{eq:I}
  I_a(k,\omega) = \frac{3\Theta_a^{3/2}}{4t} \int_0^\infty \frac{\ln\left|\frac{(t^2+2tu)^2 - (\beta\hbar\omega)^2}{(t^2-2tu)^2-(\beta\hbar\omega)^2}\right|}{\exp(u^2 - \beta\mu_a) + 1} \,u\idif u  ,
\end{multline}
where $\Theta_a=k_BT/E_{Fa}$ is the degeneracy parameter, $E_{Fa}=\hbar^2(3\pi^2n_a)^{2/3}/2m_a$ is the Fermi energy, $\mu_a$ is the chemical potential, $t^2 = \hbar^2k^2\beta/2m_a = \deBroglie_a^2 k^2$, and $\Lambda_a$ is the thermal de Broglie wavelength divided by $2\pi$.

The dynamic response functions relate to the dynamic static structure factors through the fluctuation-dissipation theorem\cite{IchimaruBook,QSCPSBook,ReinholzAPF2005}
\begin{equation}
  \label{eq:fdt}
  S_{ab}(k, \omega) = -\frac{\hbar}{2\pi} \coth(\beta\hbar\omega/2) \Im \chi_{ab}(k, \omega)
\end{equation}
from which the static structure factors are obtained as the integral over frequencies
\begin{equation}
  \label{eq:Sk-from-Skw}
  S_{ab}(k) = \frac1{\sqrt{n_an_b}} \int_{-\infty}^\infty S_{ab}(k,\omega) \idif\omega
\end{equation}
A convenient expression of this relationship is as a sum over residues
\begin{equation}
  \label{eq:matsum}
  S_{ab}(k) = -\frac{k_BT}{\sqrt{n_an_b}} \sum_{l=-\infty}^\infty \chi_{ab}(k, i\omega_l)
\end{equation}
where $\omega_l=2\pi l k_BT/\hbar$ are the Matsubara frequencies\cite{TanakaJPSJ1986}.
As will be shown below, this summation needs only to be carried out for a jellium-like response function, so convergence may be accelerated using the same technique employed by Tanaka and Ichimaru, see Eqs.~(27)-(31) of Ref.~\cite{TanakaJPSJ1986}.

At dense plasma conditions, the electron de Broglie wavelength can be of similar order as the relevant density fluctuation wavelengths, while the ion de Broglie wavelength is smaller by a factor $\sqrt{m_e/m_I}$.
This allows for considerable simplifications and an important connection to the quantum Ornstein-Zernike theory.
Taking $\deBroglie_I k\ll 1$ and $\beta\mu_I\ll 0$, the ion free-particle susceptibility for imaginary frequencies is
\begin{equation}
  \chi^0_I(k,i\omega_l) = \begin{cases}
    -\beta n_I^0 + \bigO(\deBroglie_I^2k^2) & l=0 \\
    -\beta n_I^0 \frac{\deBroglie_I^2k^2}{2\pi^2l^2} + \bigO(\deBroglie_I^4k^4l^{-4}) & l\ne 0
  \end{cases}
  .
\end{equation}
When this expansion is used in Eq.~\eqref{eq:lin-resp}, one finds for $l=0$
\begin{subequations}
  \label{eq:chi-leq0}
  \begin{align}
    \chi_{II}(k,0) &= -\beta n_I^0\frac{1-\chi_e^0(k,0)U_{ee}(k,0)}{D(k,0)} \\
    \chi_{Ie}(k,0) &= \beta n_I^0\chi_e^0(k,0) \frac{U_{Ie}(k)}{D(k,0)}  \\
    \chi_{ee}(k,0) &= \chi_e^0(k,0) \frac{1 + \beta n_I^0 U_{II}(k,0)}{D(k,0)}  \\
    \begin{split}
      D(k,0) &= 1 + n_I^0 \beta U_{II}(k,0) - \chi_e^0(k,0)U_{ee}(k,0) \\
      &\quad- n_I^0\beta\chi_e^0(k,0)[U_{II}(k,0)U_{ee}(k,0) - |U_{Ie}(k,0)|^2]
    \end{split}
  \end{align}
\end{subequations}
up to terms of order $\deBroglie_I^2 k^2$.
The corresponding expansion for $l\ne 0$ produces
\begin{subequations}
  \label{eq:chi-lne0}
  \begin{align}
    \chi_{II}(k,i\omega_l) &= -\beta n_I^0 \frac{\deBroglie_I^2k^2}{2\pi^2l^2} \\
    \chi_{Ie}(k,i\omega_l) &= \beta n_I^0\chi_e(k,i\omega_l) U_{Ie}(k,i\omega_l)\frac{\deBroglie_I^2k^2}{2\pi^2l^2} \\
    \begin{split}
      \chi_{ee}(k,i\omega_l) &= \chi_e(k,i\omega_l) \\ &\quad - \beta n_I^0 [\chi_e(k,i\omega_l)U_{Ie}(k,i\omega_l)]^2\frac{\deBroglie_I^2k^2}{2\pi^2l^2} 
    \end{split}
                                                         \label{eq:chi-ee-dyn}
  \end{align}
\end{subequations}
up to terms of order $\deBroglie_I^4 k^4 l^{-4}$.
In Eq.~\ref{eq:chi-lne0} we have defined
\begin{equation}
  \label{eq:chi-jellium}
  \chi_e(k,\omega) = \frac{\chi_e^0(k,\omega)}{1 - \chi_e^0(k,\omega)U_{ee}(k,\omega)}
\end{equation}
which is similar in form to the response function of jellium except that the polarization potential here should involve the local field correction appropriate for a TCP.

\begin{figure}
  \centering
  \includegraphics[width=\linewidth]{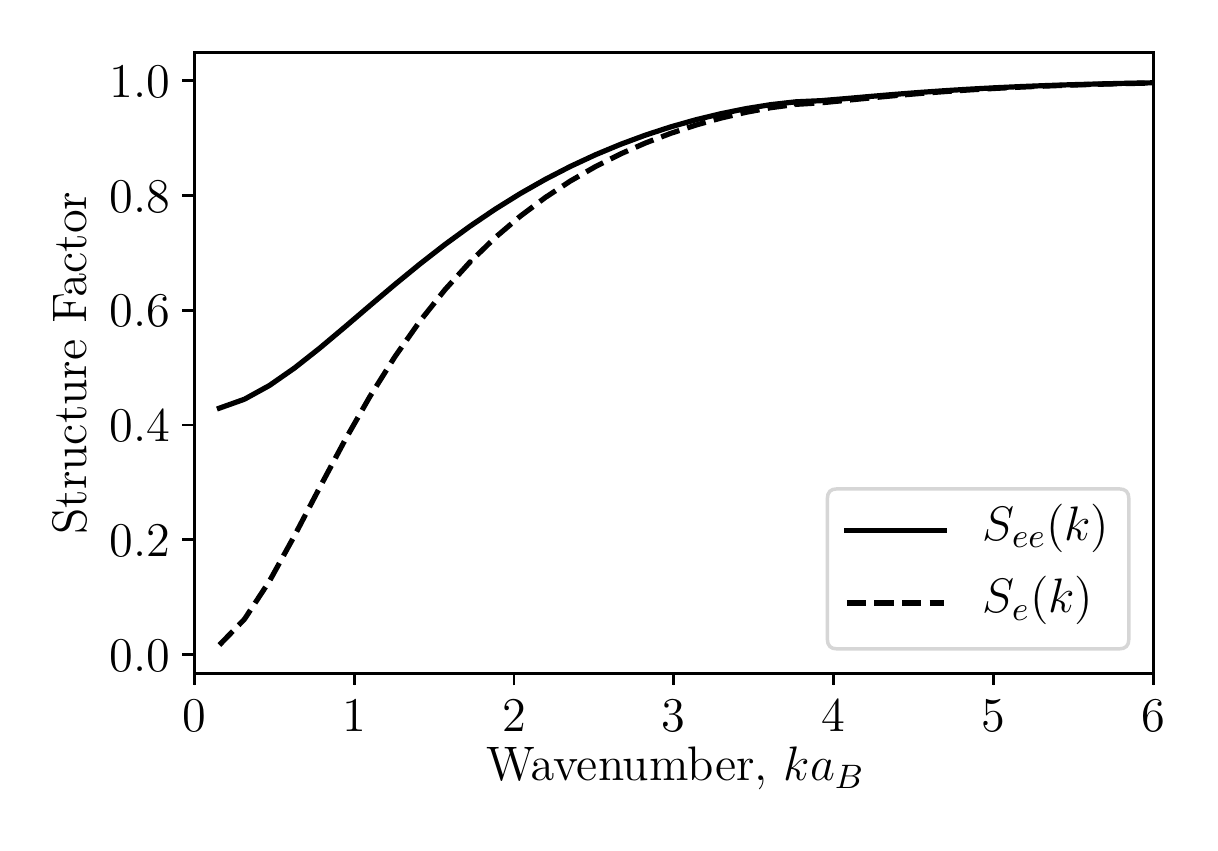}
  \caption{Electron static structre factors for hydrogen at 50eV and 2.7g/cc. The solid line is the full electron-electron structure factor, and the dashed line is the jellium contribution [first term of Eq.~\eqref{eq:See-decomp}].}
  \label{fig:hydrogen-Sk}
\end{figure}

A classical treatment of the ions corresponds to neglecting terms of order $\Lambda_I^2k^2$ and above.
Doing so, the evaluation of Eq.~\eqref{eq:matsum} for $S_{II}(k)$ and $S_{Ie}(k)$ requires only the zero-frequency ($l=0$) contribution to $\chi_{II}$ and $\chi_{Ie}$, whereas $S_{ee}(k)$ retains an $l\ne0$ contribution from the jellium-like first term of Eq.~\eqref{eq:chi-ee-dyn}
\begin{subequations}
  \begin{align}
    S_{II}(k) &= \frac{1 - \chi_e^0(k,0)U_{ee}(k,0)}{D(k,0)} \label{eq:Sii} \\
    S_{Ie}(k) &= \frac{\chi_e^0(k,0)U_{Ie}(k,0)}{\bar Z^\frac12 D(k,0)} \label{eq:Sie} \\
    \begin{split}
      S_{ee}(k) &= - \frac{1}{\beta \bar n_e^0} \sum_{l\ne0}\chi_{e}(k,i\omega_l) \\
      &\qquad- \frac{\chi_e^0(k,0)}{\beta \bar n_e^0}\frac{1 + \beta n_I^0 U_{II}(k,0)}{D(k,0)}
      \label{eq:See}
    \end{split}
  \end{align}
\end{subequations}
In their static limit, the polarization potentials are synonymous with the OZ direct correlation functions\cite{IchimaruBook,DaligaultPRE2009}
\begin{equation}
  U_{ab}(k,0) = -k_BT C_{ab}(k)
\end{equation}
and it is easy to see that in fact Eqs.~\eqref{eq:Sii} and \eqref{eq:Sie} are just the QOZ relations, Eqs.~\eqref{eq:qoz}.
For the electron-electron structure factor, a more physically illuminating formula can be written by introducing the jellium-like static structure factor,
\begin{equation}
  \label{eq:Se}
  S_e(k) = -\frac1{\beta \bar n_e^0}\sum_{l=-\infty}^\infty\chi_e(k,i\omega_l)
\end{equation}
in terms of which
\begin{equation}
  \label{eq:See-decomp}
  S_{ee}(k) = S_e(k) + \frac{\chi_e(k,0)}{\beta \bar n_e^0} - \frac{\chi_e^0(k,0)}{\beta\bar n_e^0}\frac{1 - n_I^0C_{II}(k)}{D(k,0)}
  .
\end{equation}
The first term in $S_{ee}(k)$ is just the jellium structure factor, the second term removes the jellium zero-frequency response, and the third adds back in the TCP zero-frequency response, which accounts for correlations between the electrons induced by their attraction to the ions.
The ionic correction is substantial, as shown in Fig.~\ref{fig:hydrogen-Sk}, especially at long wavelengths.
Ion correlations lift the jellium-like $S_e(k)\to 0$ behavior to a finite value as $k\to0$, which is necessary to satisfy the charge-density sum rule\cite{MartinRMP1988}.
This new expression Eq.~\eqref{eq:See-decomp} for the electron-electron static structure factor is the main result of this paper, from which other useful quantities describing electron-electron correlations can be derived.

A point of practical interest is that one can obtain accurate predictions for the static structure factors without the need for dynamic local field corrections, despite their apparent need in Eq.~\eqref{eq:matsum}.
Of the three structure factors, only $S_{ee}(k)$ involves dynamic local field corrections, and even then only in the calculation of its jellium-like part, $S_e(k)$.
Recent advances in computing the dynamic structure factor of jellium\cite{DornheimPRL2018} suggest that at high electron densities ($\bar n_e^0 \gtrsim 10^{21} \mathrm{cm}^{-3}$), the dynamic local field correction can be replaced by its static (zero-frequency) with little error in the dynamic structure factor and thus also the static structure factor, viz. Eq.~\eqref{eq:Sk-from-Skw}.
Even though the present case concerns the electron-electron dynamic local field corrections for a TCP (not jellium), we take it as a reasonable approximation that a similar result should hold here.
The results shown in Sec.~\ref{sec:results} all make use of a static electron-electron local field correction.
Approximate dynamic response is still included through the free-particle response functions, $\chi_e^0(k,i\omega_l)$, in Eqs.~\eqref{eq:chi-jellium} and~\eqref{eq:See}.

One way in which the theory could be refined concerns self-consistency.
Namely, the formulas derived in this section assume the electron-electron direct correlation function $C_{ee}$ is given.
In the practical calculations shown in Sec.~\ref{sec:results}, the jellium approximation for $C_{ee}$ is used, but clearly the resulting $S_{ee}$ will differ from that of jellium due to the second term of Eq.~\eqref{eq:See} which couples to the ions.
One could imagine constructing a self-consistent closure for $C_{ee}$ in which one starts with the jellium approximation and refines according to the resultant $S_{ee}$.
However, it is unclear how to produce an independent closure for $C_{ee}$ in terms of $S_{ee}$ or if corrections beyond the jellium approximation would make any practical difference in the resulting static structure factors.
Since $C_{ee}$ is intimately connected electron-electron exchange-correlation potential\cite{ChiharaJPCM1984}, this is an important question to resolve if the present results are to be applied to the development of new exchange-correlation or self-energy functionals.

\subsection{Pair Distribution Function and Mean-Force Potential}
\label{sec:gr-pmf}

The TCP pair distribution functions are related to the static structure factors by
\begin{equation}
  g_{ab}(r) = 1 + \frac{1}{\sqrt{n_an_b}}\int\left[S_{ab}(k) - \delta_{ab}\right]e^{i\vec k\cdot\vec r}\frac{\idif[3]k}{8\pi^3}
\end{equation}
The pair distribution functions may be used to construct potentials of mean force using Percus's theorem\cite{PercusPRL1962,ChiharaPTP1973a,AntaPRB2000}.
The theorem states that if a particle of species $a$ is inserted into the plasma at the origin, then the resulting density profile of species $b$ is given by
\begin{equation}
  \label{eq:percus}
  n_b(r | v_{ab}) = n_b g_{ab}(r)
\end{equation}
where the notation emphasizes that $n_b(r)$ is a functional of the ``external'' potential $v_{ab}(r)$.
The potential of mean force, $v^\mf_{ab}(r)$, is introduced by constructing an auxiliary system of non-interacting particles.
One then asks what external potential applied to the noninteracting system would induce the same density profile in species $b$ that is obtained when the interacting system is acted on by the external potential $v_{ab}(r)$.
This potential is the potential of mean force, and the above statement is expressed mathematically as 
\begin{equation}
  n^0_b(r|v^\mf_{ab}) = n_b(r | v_{ab})
\end{equation}
where the superscript ``0'' denotes the density profile of the non-interacting system.

An explicit formula for $v^\mf_{ab}(r)$ follows from the identity relating the chemical potential and intrinsic Helmholtz free energy $F$ of an inhomogeneous system exposed to an external potential $\phi_b(\vec r)$\cite{HansenMacDonald4e}
\begin{equation}
  \frac{\delta F}{\delta n_b(\vec r)} + \phi_{b}(\vec r) - \mu_b = 0
\end{equation}
This identity is applied separately to the interacting system exposed to $\phi_b = v_{ab}$ and to the non-interacting system exposed to $\phi_b = v^\mf_{ab}$.
Equating the two gives
\begin{equation}
  v^\mf_{ab}(\vec r) = v_{ab}(\vec r) + \frac{\delta F^{\mathrm{ex}}}{\delta n_b(\vec r)} - \mu_b^{\mathrm{ex}}
\end{equation}
where $F^{\mathrm{ex}}$ and $\mu_b^{\mathrm{ex}}$ are the non-ideal parts of intrinsic free energy and chemical potential.
The excess intrinsic free energy may be developed in a functional Taylor series about the densities of the uniform system, $n_s^0 = n_s(\vec r|v_{as})|_{v_{as}=0}$, which, after making the identifications
%
%
\begin{align}
  & \left.\frac{\delta F^{\mathrm{ex}}}{\delta n_b(\vec r)}\right|_{v_{ab}=0} = \mu_b^\mathrm{ex}\\
  & \left.\frac{\delta^2 F^{\mathrm{ex}}}{\delta n_b(\vec r) \delta n_s(\vec r')}\right|_{\substack{v_{ab}=0\\v_{as}=0}} = -\beta^{-1} C_{bs}(\vec r - \vec r') \\
  & n_s(\vec r|v_{as}) = n_s^0 g_{as}(\vec r)
\end{align}
obtains for the mean-force potential\cite{AntaPRB2000}
\begin{equation}
  \label{eq:vmf}
  v^\mf_{ab} = v_{ab} - \beta^{-1} \sum_{s=I,e} n_s (g_{sb} - 1)\star C_{as} + \beta^{-1}B_{ab}
\end{equation}
where the star denotes convolution and $B_{ab}(r)$ is the bridge function containing third- and higher-order functional derivatives of $F^{\mathrm{ex}}$.
We treat the ion-ion bridge function using the variational modified hypernetted chain approximation\cite{RosenfeldJSP1986} and neglect the electron-ion and electron-electron bridge functions, for which good approximations are not known, but should only be important when the conduction electrons are very strongly correlated.

Calculations of $v^\mf_{II}$ and $v_{Ie}^\mf$ within the present TCP model have already been applied to problems of diffusive transport in dense plasmas\cite{DaligaultPRL2016,StarrettHEDP2017,StarrettPOP2018}.
Here, we compute $g_{ee}$ and $v^\mf_{ee}$ as well.
However before presenting results, we first address an important conceptual point regarding the application of Percus's theorem to electron-electron correlations.

The application of Percus's theorem to the calculation of $v_{ee}^\mf$ introduces a semiclassical approximation.
This is because the procedure of placing a test electron at rest at the origin violates Heisenberg's uncertainty principle, since the test electron's position and momentum would be simultaneously known with perfect certainty\cite{LouisJNCS2002}.
This means that the potential of mean force computed using Percus's theorem represents a semiclassical calculation.
Since $r/\Lambda_e$ is the expansion parameter in semiclassical treatments of pair correlations in quantum gases~\cite{UhlenbeckPR1932,KirkwoodPR1933}, the validity of Eq.~\eqref{eq:vmf} for $v_{ee}^\mf$ is not guaranteed at length scales smaller than $\deBroglie_e$.
If the plasma temperature is given in electron volts, this means that $v^\mf_{ee}$ should be accurate for $r/a_B \gtrsim 5.2 T^{-\frac12}$, where $a_B$ is the Bohr radius.
As will be shown in Sec.~\ref{sec:results}, the range of $v_{ee}^\mf$ for solid density plasmas is typically on the order of a few Bohr.
For hot dense plasmas with temperatures on the order of hundreds of $\mathrm{eV}$, the disrespect of the uncertainty principle should only affect the potential at very short length scales where $v_{ee}^\mf$ differs little from the Coulomb potential.

\section{Results}
\label{sec:results}

\subsection{Comparison with First-Principles Simulations}
\label{sec:pimc-compare}

Electron-electron correlation physics in warm and hot dense plasmas is difficult to assess by first-principles means.
In particular, while Kohn-Sham molecular dynamics (QMD) simulation is a useful methodology for benchmarking theoretical models of ionic correlations, the physics of electron correlation exists only in the choice of exchange-correlation functional used to compute the electron density.
QMD is thus not a useful means of assessing the present model's accuracy.
Path integral Monte Carlo (PIMC) methods, however, offer a high-fidelity description of electron-electron correlations.
A challenge in connecting the present model with PIMC is that PIMC studies in general treat a plasma as a system of nuclei and electrons (both bound and free) whereas the AA-TCP model assigns some fraction of the electron density to the nucleus to construct ions.
To compare with PIMC results for $g_{ee}(r)$, we are thus limited to materials at high enough temperatures and densities that there are no electrons bound to the nucleus.

\begin{figure}
  \centering
  \includegraphics[width=\linewidth]{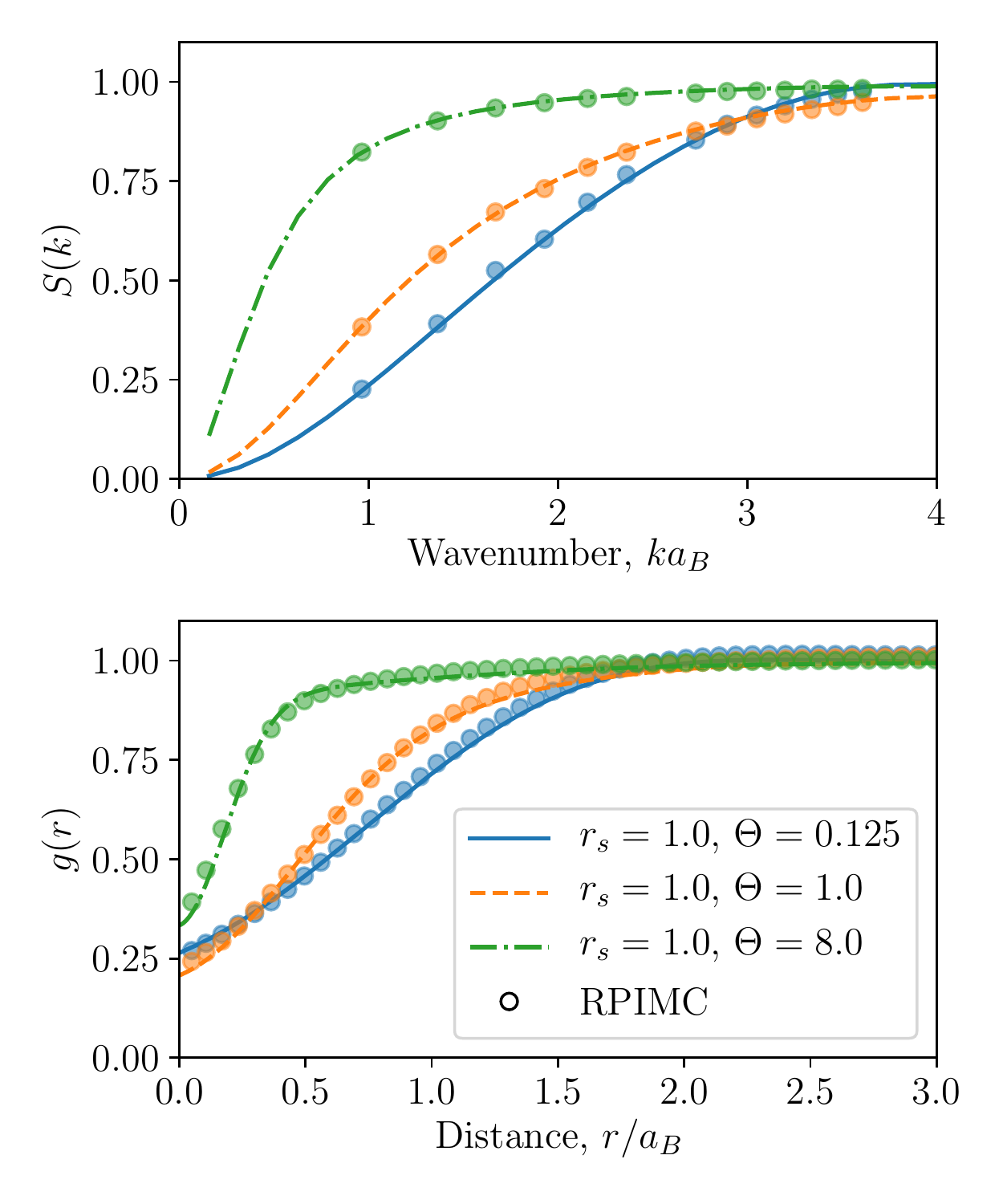}
  \caption{Pair distribution functions (upper) and static structure factors (lower) of jellium. Lines are the jellium model used in AA-TCP. Circles are restricted-PIMC results by Brown et al.\cite{BrownPRL2013}}
  \label{fig:jellium-gr-vs-pimc}
\end{figure}

The simplest such ``material'' is the jellium model.
It is important even in the present context, since the jellium structure factor appears a term in the electron-electron structure factor, as derived in Eq.~\eqref{eq:See-decomp}.
Fig.~\ref{fig:jellium-gr-vs-pimc} affirms that the jellium contribution to electronic correlations is accurately treated in the AA-TCP model, as compared with restricted-PIMC simulations by Brown et al.~\cite{BrownPRL2013}.
Comparisons are shown for electron densities corresponding to $r_s = 1$, where $r_s = a_e/a_B$ and $a_e = (4\pi \bar n_e^0/3)^{-\frac13}$, which is typical of near-solid density plasmas.

Turning now to real matter, Fig.~\ref{fig:deuterium-gr-dos-vs-pimc}
compares the pair distribution functions of the AA-TCP model with
those computed from PIMC by Militzer for warm dense
deuterium\cite{MilitzerThesis}.
Due to computational constraints on the number of particles at the time, the PIMC pair distribution functions do not asymptote to unity at large separation{, instead taking values up to a few percent above or below unity.
  To best connect with the AA-TCP model, which occurs in the thermodynamic limit, the PIMC pair distribution
  functions have been rescaled $g_{ab}(r) \to g_{ab}(r) / g_{ab}(r_{\max})$, where $r_{\max}$ is the largest tabulated separation.
  Furthermore, since the PIMC electrons have spin, the overall electron-electron pair distribution has been constructed as the mean of the two spin orientations\footnote{
    In general, the parallel and antiparallel spin contributions to $g_{ee}(r)$ should be weighed not by $\frac12$ but by $\frac12 \frac{N-2}{N-1}$ and $\frac12 \frac{N}{N-1}$, respectively ($N$ being the number of electrons).  In neglecting this, we incur errors of order $1/N$ in forming the PIMC $g_{ee}(r)$, which is consistent with our treatment in processing the PIMC data as if in the thermodynamic limit.
  }.

  The conditions of Fig.~\ref{fig:deuterium-gr-dos-vs-pimc} represent a stringent test of the AA-TCP model because at the temperature shown, $10.8\mathrm{eV}$, the electronic structure of deuterium is sensitive to the density.
  It is observed that the AA-TCP model systematically underestimates the depth of the electron-electron correlation hole, and that the disagreement is greater at lower density.
  The tendency for the AA-TCP model to underestimate the degree of electron-electron correlation can be qualitatively understood by inspecting the electronic density of states (DOS) of the average-atom model.
  This DOS is obtained in an ion-sphere average-atom calculation as an intermediate step to constructing the TCP (See the Appendix and Ref.~\cite{StarrettPRE2013} for the distinction between the two).
  In contrast, the conduction electrons of the TCP should be thought of as being nearly free with an ideal ($\propto\sqrt E$) DOS.

  The ion-sphere average-atom DOS exhibits a resonance-like feature in the low-energy part of the continuum, corresponding to electrons which are not bound to the nucleus but still strongly interact with it.
  This feature in the DOS is sharpest at the lower densities shown, coinciding with the conditions where AA-TCP model is in greatest disagreement with PIMC.
  With increasing density, the non-free feature in the DOS broadens and shifts further out into the continuum, the electrons are less strongly correlated, and the AA-TCP model is in good agreement with PIMC.
  The exclusion principle offers a simple, if loose, explanation: at higher density (smaller ion-sphere), the continuum electrons' spatial distribution compresses, so their energy (momentum) distribution must broaden.

  The onset of strong electron correlation features in the the DOS is symptomatic of the breakdown of the TCP concept, rather than our theory for the electron-electron correlations specifically.
  This is because the presence of barely-free electrons makes it difficult to unambiguously define an ``ion'' as a distinct entity.
  Indeed, the appearance of these long-lived resonance-like states renders all three AA-TCP pair distribution functions inaccurate compared with PIMC, not just $g_{ee}(r)$.
  The Appendix gives a more quantitative discussion of this breakdown in terms of the accuracy of the AA-TCP electron-ion closure.
  \begin{figure*}
    \centering
    \includegraphics[width=\linewidth]{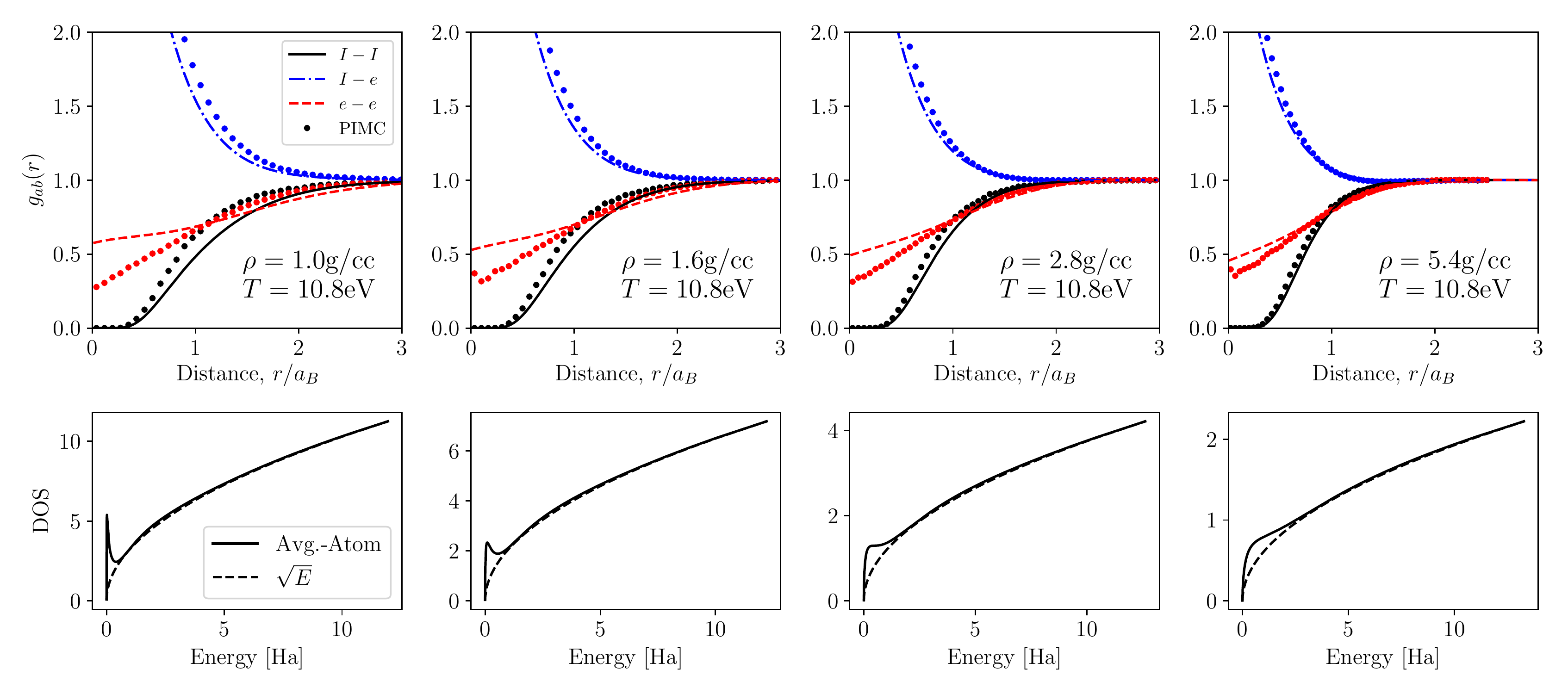}
    \caption{Upper: pair distribution functions of warm dense deuterium. Black solid, blue dash-dotted, and red dashed lines are the TCP model $g_{II}(r)$, $g_{Ie}(r)$, and $g_{ee}(r)$ respectively. Circles are PIMC results by Militzer\cite{MilitzerThesis}.
      Lower: average-atom electronic density of states. Solid lines are the DOS of the average-atom model. Dashed lines are the free-electron DOS assumed in constructing the TCP.}
    \label{fig:deuterium-gr-dos-vs-pimc}
  \end{figure*}

  Available PIMC results also allow for verification of the high-temperature limiting behavior of $g_{ee}(r)$ in higher-$Z$ materials.
  Figure~\ref{fig:solidAl-gee-highT} compares the electron-electron pair distribution functions of solid-density aluminum (2.7 g/cm$^3$) with the PIMC results obtained by Driver et al.\cite{DriverPRE2018}.
  At the temperatures shown, both the PIMC simulations and the AA-TCP model predict the aluminum is fully ionized, so direct comparisons between the two methods are possible.
  All departures from the classical ideal $g_{ee}(r)=1$ behavior are confined to distances less than about one Bohr, which is much smaller than the relevant interaction range, the Debye length.
  The AA-TCP model and PIMC results are in good agreement with the jellium treatment, in which the ionic correlations are absent.
  The electron subsystem of the TCP is thus effectively decoupled from the ions.
  Additionally, neither the PIMC results nor the TCP model differ much from the analytic form for a nearly-classical ideal Fermi gas, for which
  \begin{equation}
    g_e^0(r) \approx 1 - \frac12 \exp\left(-\frac12 \frac{r^2}{\Lambda_e^2}\right)
  \end{equation}
  and all departures from the classical $g_{ee}=1$ behavior are due to exchange\cite{DiesendorfJMP1968}.
  The PIMC results do exhibit some slight fluctuation in regions where the theoretical models predict $g_{ee}$ to be unity.
  These result from a not-quite-exact cancellation of the parallel- and antiparallel-spin channels, which are resolved in PIMC but absent from the TCP treatment.
  It is unclear whether this is a physical effect or a consequence of simple statistical variability intrinsic to the PIMC method.
  Even if these spin-dependent fluctuations are physical, in this high-temperature limit they are confined to relatively short length scales (Bohr versus Debye lengths) and are unlikely to make any difference in practical applications.
  \begin{figure}
    \centering
    \includegraphics[width=\linewidth]{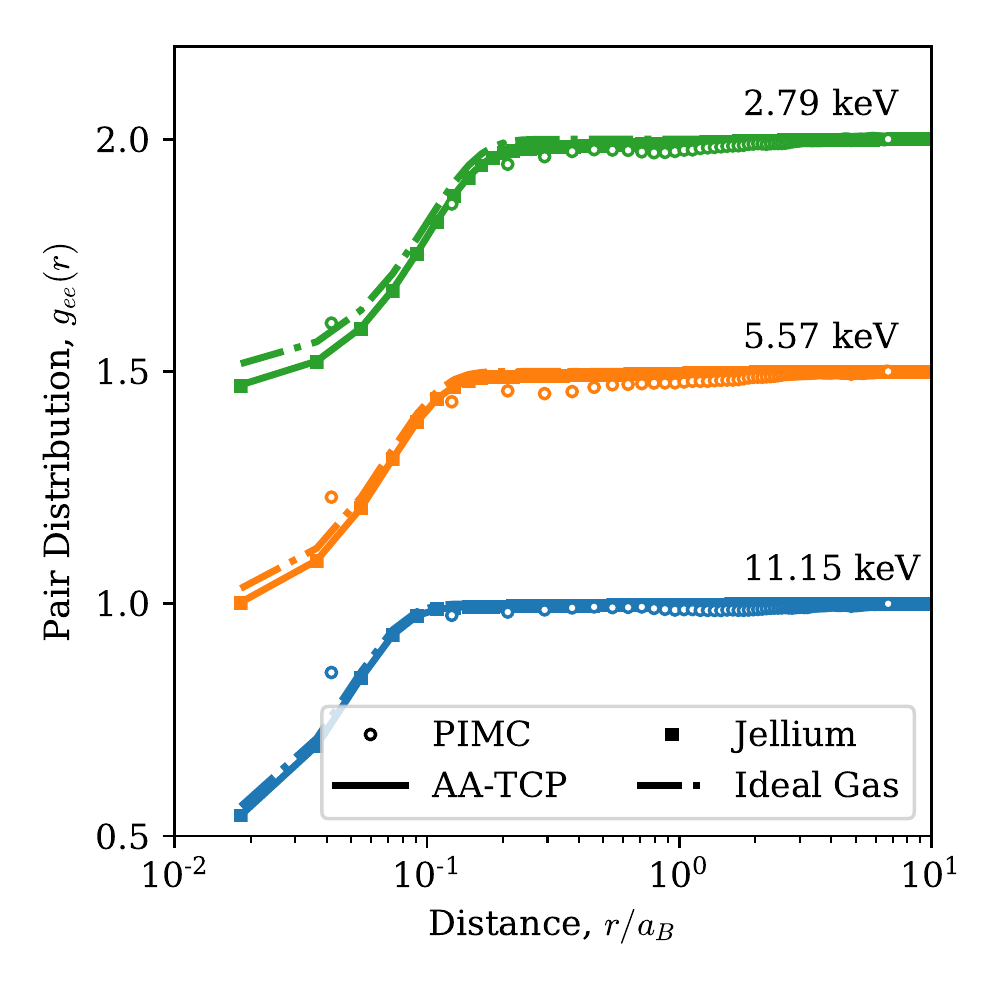}
    \caption{Electron-electron pair distribution functions of solid-density aluminum. Circles are spin-averaged PIMC results by Driver et al.\cite{DriverPRE2018}. Solid lines are the AA-TCP model. Squares are the jellium part of the AA-TCP model. Dash-dotted lines are for an ideal and nearly classical Fermi gas. Each set of data is offset vertically in increments of one-half.}
    \label{fig:solidAl-gee-highT}
  \end{figure}

  \subsection{Potentials of Mean Force}
  \label{sec:vmf}

  Figure~\ref{fig:aluminum-vmf} shows the electron-electron potentials of mean force for solid-density aluminum.
  The asymptotic $e^2/r$ dependence as $r\to0$ is divided out to emphasize the screening part of the potential.
  The AA-TCP model is compared with two simplified treatments.
  The first is to treat the electron-electron correlations in the random phase approximation (RPA), corresponding to approximating the polarization potential by the bare Coulomb interaction, $G_{ee}(k,\omega) \approx 0$.
  The second limit shown is that of high temperatures, where the potential of mean force reduces to a simple screened interaction\cite{HansenMacDonald4e,ShafferPOP2017}
  \begin{equation}
    \label{eq:vmf-wc}
    v_{ee}^\mf(r) \to \frac{e^2}{r}\exp(-\kappa r)
  \end{equation}
  Here, the inverse screening length is given by $\kappa = \sqrt{\kappa_I^2 + \kappa_e^2}$, with
  $\kappa_s = \sqrt{4\pi \bar Z_s^2 e^2 \beta n_s^0}$ being the Debye wavenumber of either species.
  This limit is reached when all correlations are treated in the RPA and the dynamic electron screening is treated classically, i.e., the first term of Eq.~\eqref{eq:See} is dropped.

  At 1000eV, the aluminum is nearly fully stripped ($\bar Z=12.6$) and essentially classical ($\Theta_e = 32.9$).
  Simple exponential screening is a very good approximation to the full AA-TCP model at these conditions.
  At 100eV ($\bar Z=7.87$, $\Theta_e=4.5$), the temperature is high enough that the RPA offers a good description of the electron-electron correlations but the screening is distinctly non-exponential due to indirect correlations with the ions, which are strongly coupled due to their relatively high charge.
  At 10eV ($\bar Z=3.02$, $\Theta_e=0.853$), these indirect correlations dominate the screening at distances less than the inter-ionic spacing $a_I = 2.99 a_B$.
  This occurs because in the average-atom calculation underlying the TCP construction, the continuum electrons are correlated to the ions' bound electrons by the condition that all orbitals be mutually orthogonal.
  The RPA manages to qualitatively capture this effect since the electron-ion correlations are still being treated fully, but it is quantitatively deficient compared with the full AA-TCP treatment.
  At 10eV, it is also clear that exponential screening is a completely unsuitable description of the electron-electron mean-force potential.
  The apparent attractive feature in $v^\mf_{ee}(r)$ near $r\approx 3.5 a_B$ is an ionic structure effect, whereby the accumulation of ions at this distance induces electron correlations.
  \begin{figure}
    \centering
    \includegraphics[width=\linewidth]{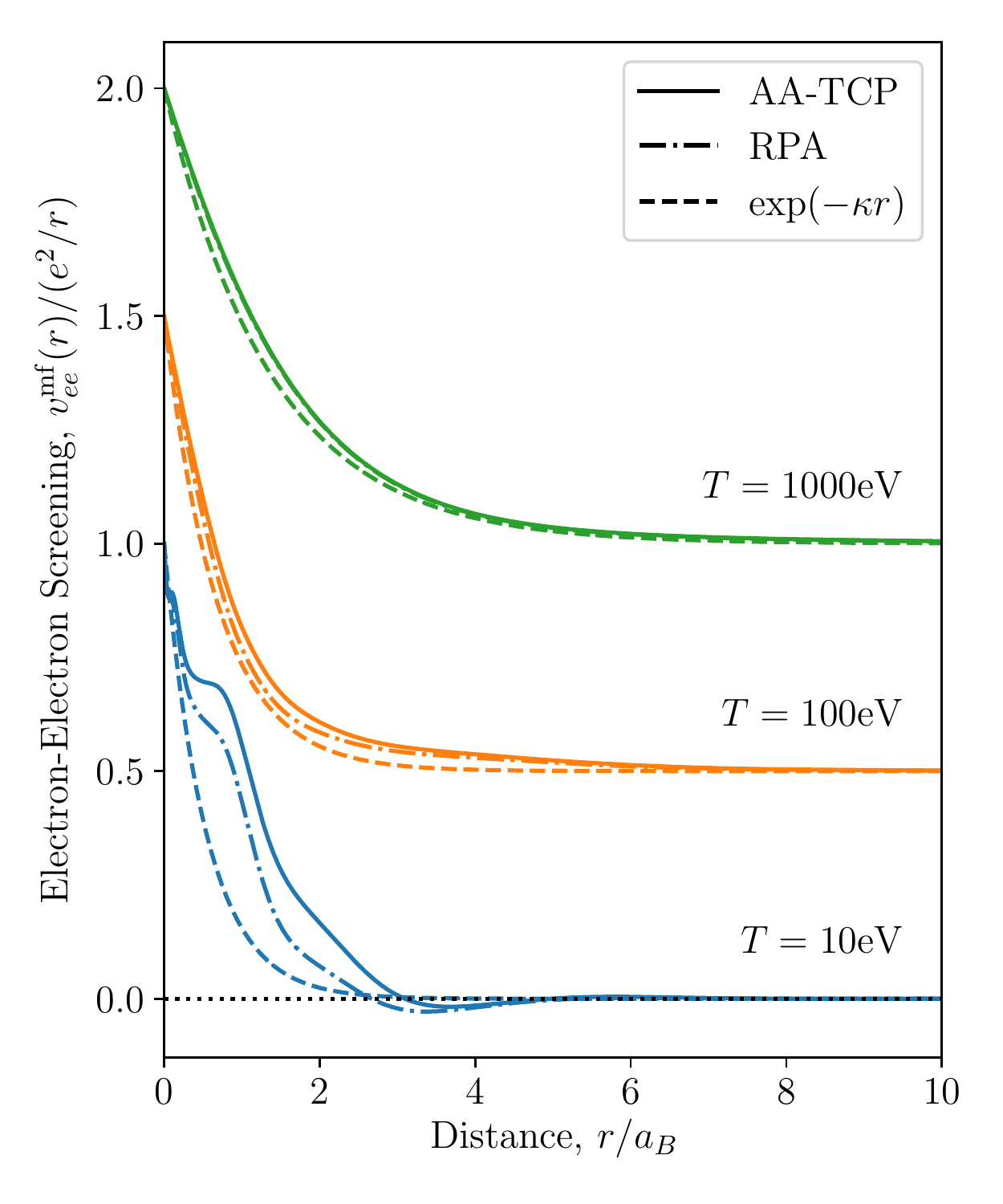}
    \caption{Screening part of the electron-electron potential of mean force for solid-density aluminum at temperatures from 10 to 1000 eV. Solid lines are the AA-TCP model. Dash-dotted lines are the AA-TCP model with $U_{ee}$ treated in the RPA. Dashed lines are the high-temperature limit, given in Eq.~\eqref{eq:vmf-wc}. Each pair of curves is offset vertically.}
    \label{fig:aluminum-vmf}
  \end{figure}

  \section{Conclusions}
  \label{sec:conc}

  We have derived a new formula for the electron-electron static structure factor that is suitable for plasmas of classical ions and quantum-mechanical electrons.
  The formula naturally completes the quantum Ornstein-Zernike relations which provide a unified description of ionic and electronic structure but which could not have been used to treat electron-electron correlations until now.
  In the present work, we have focused on plasmas with a single ion species for definiteness, but the final analytic formula for the electron-electron structure factor extends in a straightforward (if algebraically cumbersome) way to the case of multiple ion species.
  Evaluating the theory for mixtures using average-atom models should give accurate results at similar conditions as for pure plasmas, provided that molecular bonds do not form\cite{StarrettPRE2014}.}
With the static approximation for the electron-electron local field corrections, the electron-electron structure factor may easily be computed from an average atom model.
Comparison with path integral Monte Carlo results demonstrated that the resulting electron-electron pair distribution functions are accurate provided that the conduction electrons are not too strongly correlated with one another, e.g., due to the appearance of resonances.
However, such conditions represent a breakdown of the underlying concept of distinct ions and conduction electrons rather than the theory itself.

Improved knowledge of static pair correlations between the conduction electrons in dense plasma should stimulate interest in translating modern theories of electron correlation in solids to the plasma state, which is more commonly treated as a mixture of ions and free electrons rather than nuclei and electrons.
In particular, it seems natural to use our results develop new approximations in the vein of either Green's function frameworks such as $GW$ or adiabatic time-dependent density functional theory.

We have also constructed electron-electron potentials of mean force which represent an effective electron-electron interaction potential.
Comparison with the Debye-H\"uckel limit showed that the electron-electron screening can be significantly affected both by the indirect influence of strongly coupled ions as well as due to correlations induced by the orthogonality of the conduction electron states to the bound electrons.
These departures from weak-coupling behavior could significantly affect the effective binary scattering physics of the electrons and could influence the electron-electron scattering contributions to electrical and thermal conductivities of dense plasmas.
Such effects could be investigated, for example, within a mean-force Boltzmann approach\cite{BaalrudPRL2013,BaalrudPOP2019} or a dynamic-screening generalized linear response approach\cite{ReinholzPRE2012,ReinholzPRE2015}.

\appendix*

\section{Closures for the AA-TCP Model}
\label{sec:closures}

This Appendix summarizes the closures used to evaluate the AA-TCP model in this work.
The formulation and closure of the AA-TCP model is discussed at length in Ref.~\cite{StarrettPRE2013}.

The formally exact ion-ion closure is known from the theory of classical fluids~\cite{HansenMacDonald4e}
\begin{equation}
  \label{eq:ii-closure}
  \ln g_{II}(r) =  -\beta \frac{\bar Z^2}{r} + g_{II}(r) - 1 - C_{II}(r) + B_{II}(r)
\end{equation}
where $g_{II}(r) = 1 + (8\pi^3n_I^0)^{-1}\int[S_{II}(k) - 1]e^{i\vec k\cdot\vec r}\idif[3]k$ is the ion-ion pair distribution function, and $B_{II}(r)$ is the bridge function.
The bridge function here is computed in the variational modified hypernetted chain approximation\cite{RosenfeldJSP1986}.

For the ion-electron closure, we obtain $C_{Ie}$ by identifying the screening density $n_e^\scr$ in Eq.~\eqref{eq:qoz-nescr} with that from a sequence of two electronic structure calculations.
The first obtains $n_e(r)$, the density of electrons about a nucleus assuming a homogeneous plasma of identical surrounding ions.
A fraction of the electron density is assigned to the nucleus, which defines an ``ion'' through the density $n_e^{\mathrm{ion}}$.
The second electron structure calculation obtains $n_e^{\mathrm{ext}}(r)$, which is solved for in the same way as $n_e(r)$, except that the central nucleus is omitted; it is the density of electrons around the nucleus which is due to the other ions.
The screening density is then formed as $n_e^\scr = n_e - n_e^\mathrm{ext} - n_e^\mathrm{ion}$, which is the density of electrons responsible for screening an individual ion.
The screening density also determines the mean ionization $\bar Z = \int n_e^\scr(r)\idif[3]r$ and thus also the mean conduction electron density, $\bar n_e^0 = \bar Z n_I^0$.
All the electronic structure calculations performed for this work used Kohn-Sham-Mermin density functional theory with the KSDT finite-$T$ exchange-correlation functional\cite{KarasievPRL2014}.

For the electron-electron closure, we set $C_{ee}$ to be the direct correlation function of jellium with the same number density and temperature as the conduction electrons of the TCP.
The direct correlation function of jellium (or equivalently its local field corrections) have been parameterized by many authors.
Our implementation uses one by Chabrier, which includes temperature dependence\cite{ChabrierJPF1990}.
One could also interpolate the tabulated results of PIMC simulations\cite{DornheimPR2018}.

\begin{figure}
  \centering
  \includegraphics[width=\linewidth]{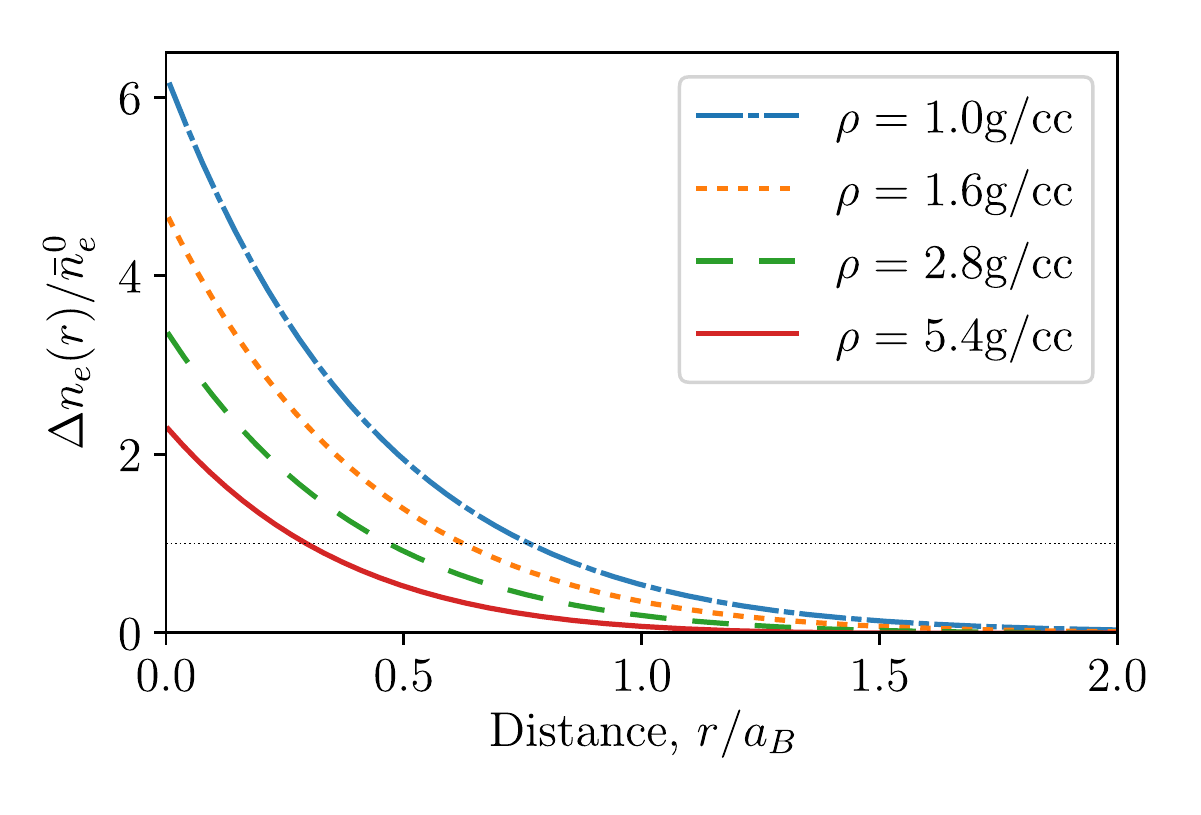}
  \caption{Perturbation in the electron density due to the ion for hydrogen at 10.8eV and the same densities shown in Fig.~\ref{fig:deuterium-gr-dos-vs-pimc}. The dotted line at unity indicates where the perturbed density is equal to the reference uniform density.}
  \label{fig:hydrogen-deltane}
\end{figure}
The electron-ion closure warrants a few additional comments, since it is closely connected with the viability of constructing a two-component plasma model from the average-atom calculation.
The closure can be expected to be accurate wherever the density profile of free electrons around an ion, $\bar n_e^0 g_{Ie}(r)$, is small compared to the mean electron density, $\bar n_e^0$.
If this is not the case, the concept of a plasma of ions and nearly free electrons breaks down.
The smallness of the perturbed electron density $\Delta n_e(r) = n_e^0 [g_{Ie}(r) - 1]$ also serves as a rough indicator for the convergence of the functional Taylor series expansion of the free energy underlying the variational formulation of the AA-TCP model, see Eq.~(28) of Ref.~\cite{StarrettPRE2013}.
In Fig.~\ref{fig:hydrogen-deltane} we plot the relative density perturbation for deuterium at the same conditions shown in Fig.~\ref{fig:deuterium-gr-dos-vs-pimc}.
At all conditions, the electron density perturbation is large near the nucleus, but this represents only a small amount of the total electron density.
The relevant figure of merit is to see how far, $r_*$, one must venture from the nucleus before the perturbation drops below unity.
The fraction of perturbed electrons within this range gives a good indication for the accuracy of the closure.
For the conditions plotted, these values are tabulated in Table~\ref{tab:perturb}, computed as
\begin{equation}
  f = \frac{ \int_0^{r_*} r^2 \Delta n_e(r) dr}{\int_0^\infty r^2\Delta n_e(r) dr}
\end{equation}
At high densities, where the AA-TCP model is in fair agreement with PIMC, only about 11\% of the perturbed electron density lies within $r_*$.
At lower densities, where the AA-TCP model is in poor agreement with PIMC, about one third of the perturbed electrons are strongly perturbed.
\begin{table}
  \centering
  \begin{tabular}{c|cccc}
    $\rho$ [g/cc] & 1.0   & 1.6   & 2.8   & 5.4 \\\hline
    $r_*$ [$a_B$] & 0.761 & 0.663 & 0.464 & 0.293 \\
    $f$           & 0.335 & 0.277 & 0.198 & 0.110
  \end{tabular}
  \caption{Electron-ion closure figures of merit for deuterium at 10.8 eV}
  \label{tab:perturb}
\end{table}

\begin{acknowledgments}
  We wish to thank Travis Sjostrom and Patrick Hollebon for useful discussions, Ethan Brown for making available the jellium PIMC data, and Burkhard Militzer for sharing the deuterium and aluminum PIMC data.
  This work was performed under the auspices of the United States Department of Energy under Contract No.~89233218CNA000001.
\end{acknowledgments}

\bibliography{refs}

\end{document}